\documentclass[prl,aps,twocolumn,superscriptaddress,notitlepage,reprint]{revtex4-1}

\pdfoutput=1

\usepackage{graphicx}
\usepackage{amsmath}
\usepackage{amssymb}
\usepackage{times}
\usepackage{color}
\usepackage{bbm}
\usepackage{bm}

\begin{document}

\title{Learning an unknown transformation via a genetic approach}

\author{Nicol\`o Spagnolo}
\email{nicolo.spagnolo@uniroma1.it}
\affiliation{Dipartimento di Fisica, Sapienza Universit\`{a} di Roma,
Piazzale Aldo Moro 5, I-00185 Roma, Italy}

\author{Enrico Maiorino}
\affiliation{Dipartimento di Fisica, Sapienza Universit\`{a} di Roma,
Piazzale Aldo Moro 5, I-00185 Roma, Italy}

\author{Chiara Vitelli}
\affiliation{Dipartimento di Fisica, Sapienza Universit\`{a} di Roma,
Piazzale Aldo Moro 5, I-00185 Roma, Italy}

\author{Marco Bentivegna}
\affiliation{Dipartimento di Fisica, Sapienza Universit\`{a} di Roma,
Piazzale Aldo Moro 5, I-00185 Roma, Italy}

\author{Andrea Crespi}
\affiliation{Istituto di Fotonica e Nanotecnologie, Consiglio
Nazionale delle Ricerche (IFN-CNR), Piazza Leonardo da Vinci, 32,
I-20133 Milano, Italy}
\affiliation{Dipartimento di Fisica, Politecnico di Milano, Piazza
Leonardo da Vinci, 32, I-20133 Milano, Italy}

\author{Roberta Ramponi}
\affiliation{Istituto di Fotonica e Nanotecnologie, Consiglio
Nazionale delle Ricerche (IFN-CNR), Piazza Leonardo da Vinci, 32,
I-20133 Milano, Italy}
\affiliation{Dipartimento di Fisica, Politecnico di Milano, Piazza
Leonardo da Vinci, 32, I-20133 Milano, Italy}

\author{Paolo Mataloni}
\affiliation{Dipartimento di Fisica, Sapienza Universit\`{a} di Roma,
Piazzale Aldo Moro 5, I-00185 Roma, Italy}

\author{Roberto Osellame}
\affiliation{Istituto di Fotonica e Nanotecnologie, Consiglio
Nazionale delle Ricerche (IFN-CNR), Piazza Leonardo da Vinci, 32,
I-20133 Milano, Italy}
\affiliation{Dipartimento di Fisica, Politecnico di Milano, Piazza
Leonardo da Vinci, 32, I-20133 Milano, Italy}

\author{Fabio Sciarrino}
\email{fabio.sciarrino@uniroma1.it}
\affiliation{Dipartimento di Fisica, Sapienza Universit\`{a} di Roma,
Piazzale Aldo Moro 5, I-00185 Roma, Italy}

\begin{abstract}
Recent developments in integrated photonics technology are opening the way to the fabrication of complex linear optical interferometers. The application of this platform is ubiquitous in quantum information science, from quantum simulation to quantum metrology, including the quest for quantum supremacy via the boson sampling problem. Within these contexts, the capability to learn efficiently the unitary operation of the implemented interferometers becomes a crucial requirement. In this letter we develop a reconstruction algorithm based on a genetic approach, which can be adopted as a tool to characterize an unknown linear optical network. We report an experimental test of the described method by performing the reconstruction of a $7$-mode interferometer implemented via the femtosecond laser writing technique. Further applications of genetic approaches can be found in other contexts, such as quantum metrology or learning unknown general Hamiltonian evolutions. 
\end{abstract}

\maketitle

\textit{Introduction.-} Linear optical networks have recently received increasing attention in the quantum regime thanks to the enhanced capability of building complex interferometers made possible by integrated photonics. This experimental achievement opened new perspectives in the adoption of linear optical networks for different quantum tasks, including quantum walks and quantum simulation \cite{Perets2008,Broome2010,Peru2010,Schreiber2011,Owens2011,Kitagawa2012,Schreiber2012,Sans2012,Crespi2013,Pits16}, quantum phase estimation \cite{Spagnolo2012,Chab14,Ciam15}, as well as the experimental implementation of the Boson Sampling problem \cite{Broome2013,Spring2013,Till2012,Crespi2012,Spagnolo2013,Spagnolo2013a,Carolan2013a,Bent15}.
\begin{figure*}[ht!]
\centering
\includegraphics[width=0.99\textwidth]{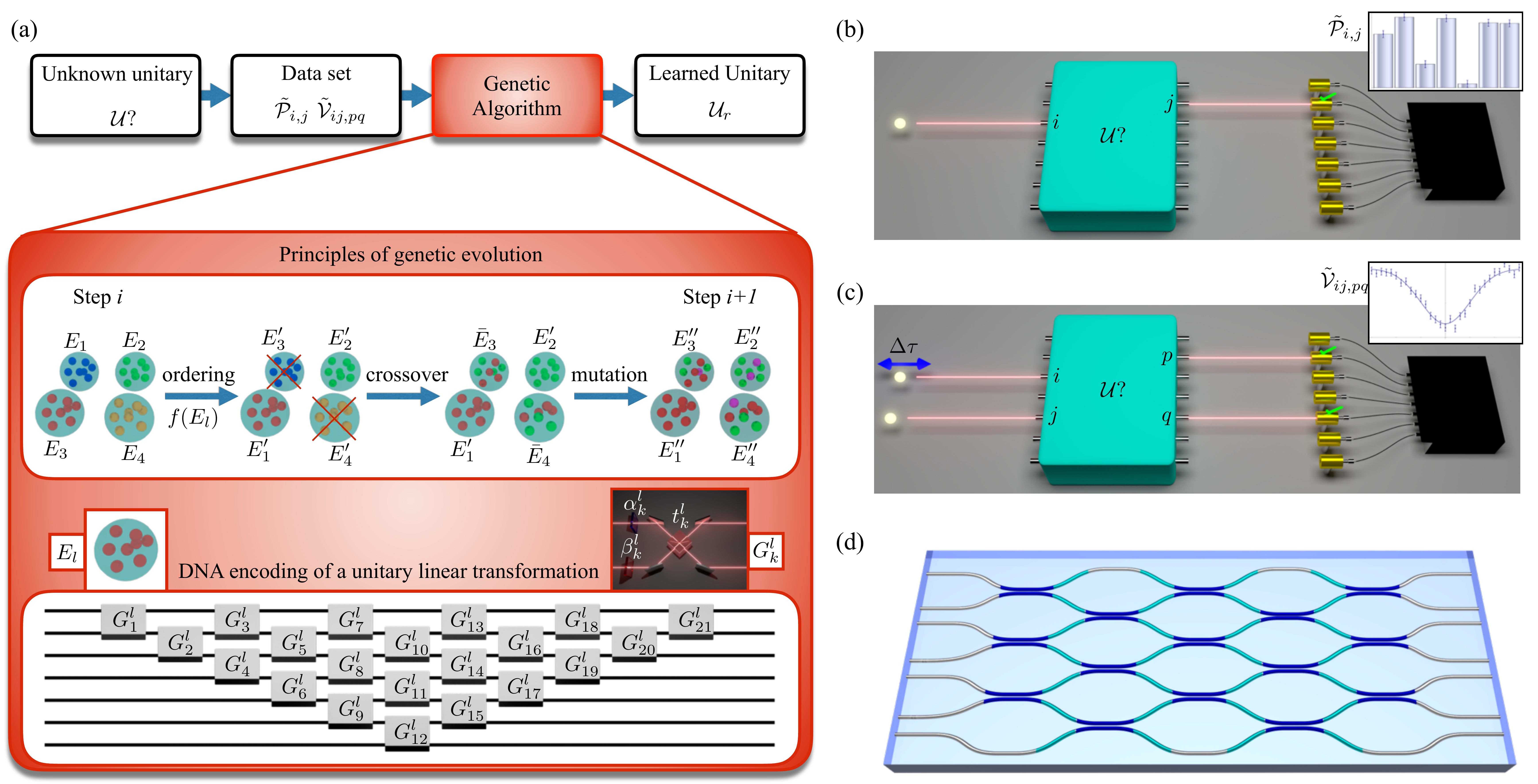}
\caption{(a) Learning an unknown linear unitary transformation via a genetic approach. The input data set, measured from the unknown transformation, are processed by an algorithm based on the principles of biological systems. The unitary transformation is decomposed in elementary units, i.e. the genes $G^l_k$ composing its DNA: beam-splitters (BSs) with transmittivity $t^l_k$ and phase-shifts (PSs) $\alpha^l_k,\beta^l_k$. Crossover and mutation mechanism rule the evolution for each step of the algorithm. (b) Schematic view of single-photon measurements corresponding to data set $\tilde{\mathcal{P}}_{i,j}$. (c) Schematic view of two-photon measurements corresponding to data set $\tilde{\mathcal{V}}_{ij,pq}$. (d) Internal structure of the actual characterized $m=7$ integrated linear interferometer. Blue regions indicate directional couplers, that is, integrated versions of beam-splitters, while cyan regions indicated phase shifts, introduced by modifying the optical path of the waveguides.}
\label{fig:schematriangolo}
\end{figure*}
Within these contexts, it becomes a crucial task to learn the action of a linear process. On one side, the capability of efficiently reconstructing an unknown transformation provides an analysis tool for integrated devices. Indeed, it allows to verify the quality of the fabrication by checking the adherence of an implemented transformation with the desired one. Conversely, on the fundamental side precise knowledge of the unitary process is required in several tasks to perform accurate tests on the experimental data. For instance, this holds in the case of Boson Sampling validation, where the adoption of statistical tests may require knowledge of the implemented unitary transformation \cite{Spagnolo2013a,Carolan2013a}. Furthermore, the task of learning an unknown transformation can be in principle embedded into a larger class of problems, whose objective is to learn physical evolutions from training sets of data \cite{Gran12}.

While initial efforts have been dedicated to the characterization of generic quantum processes \cite{Altepeter2003,OBrien2004,Rohde2005,Mohseni2006,Mohseni2008,Lobi2008,Bongio2010,Ferr2012}, different methods have been specifically adopted and tested to reconstruct an unknown linear transformation $\mathcal{U}$ \cite{Peru2011,Obrien12,Kesh2013,Dhan15,Till15}. Most of these approaches rely on single-photon and two-photon measurements. Intuitively, single-photon states can be used to obtain information on the square moduli of the unitary matrix, while two-photon interference provides knowledge of the complex phases of the elements of $\mathcal{U}$. Different data analysis approaches have been proposed and adopted to convert the raw measured data in an estimated unitary $\overline{\mathcal{U}}$, exploiting conventional numerical minimization techniques \cite{Peru2011} or by analytically inverting the relations between experimentally measured data and the elements of $\mathcal{U}$ \cite{Obrien12}. Other methods exploit classical light as input in the interferometer \cite{Kesh2013}. In this case, knowledge on the moduli is obtained by sending classical light on a single input, while knowledge on the phases is obtained by sending light on pairs of input modes and by measuring the interference fringes in the output intensities as a function of the relative phase.

In this letter we discuss and test experimentally an approach for the reconstruction of linear optical interferometers based on the class of genetic algorithms \cite{Whitley1994,Mitchell1996,Schm01}. The latter is a general method that exploits the principles of natural selection in the evolution of a biological system, and has found application to find the solution to optimization and search problems in several fields, including first applications in quantum information tasks \cite{Bang14,LasH15}. We first discuss the general principles of operations of the broad class of genetic algorithms. Then, we show how to adapt these principles of operations to the specific case of linear optical networks tomography. Finally, we test experimentally the genetic algorithm by performing the reconstruction of a $m=7$ modes integrated interferometer built by the femtosecond laser-writing technique \cite{gattass2008flm,dellavalle2009mpd}. 

\textit{Genetic reconstruction algorithm for unitary transformations.-} Genetic algorithms are a broad class of algorithms inspired by the natural evolution of biological systems, which evolve following the principle of natural selection \cite{Whitley1994,Mitchell1996,Schm01}. This principle can be briefly described as follows: within an ecosystem, individuals struggling for survival coexist within the same population. Genetically fittest individuals, e.g. individuals with highest adaption to environmental variables, are more likely to survive and reproduce. The fitness of an individual is determined by its genetic signature, the DNA, which is composed by a set of genes representing its fundamental units. Two individuals generate the offspring that inherits a combination of the genes belonging to both the parents by means of reproduction. Thus, at variance with the DNA as a whole, a single gene is or is not inherited but cannot be partially inherited. If the combination of inherited genes determines a better fitness than the parents' one, the son will have higher survival probability. Since weaker individuals are more unlikely to survive, fittest genes are more likely to spread over the population and, consequently, a gradual improvement of the average fitness of the population is expected. The set of genes belonging to all the individuals of a given population is called genetic pool. The described evolution, however, would be destined to reach a local maximum since the evolved genetic pool would be composed of just a subset of the initial genetic pool. Indeed, the mechanism of reproduction, as said, allows for the recombination of existing genes, but not for the creation of new ones. This would imply that the maximum possible fitness reached by any individual of the population would strongly depend on the initial genetic pool. Hence, it is crucial to consider in this model also the mechanism of mutation \cite{Whitley1994,Mitchell1996,Schm01}. The latter is a rare event that manifests when an inherited gene changes its form, i.e. mutates, in a random fashion. This mutated gene would likely not be present in any of the parents' DNA and could possibly provide new advantageous features causing the increase of the individual's probability of survival and reproduction. This will allow the mutated gene to spread over the population by reproduction, increasing the maximum fitness achievable within the given genetic pool. By adopting the mechanism of mutation, the evolution is no longer limited by the initial conditions and is thus more effective.

The principles of genetic evolution can be applied to learning an unknown unitary linear transformation (Fig. \ref{fig:schematriangolo}a). The goal is to find the unitary matrix $\mathcal{U}_r$  whose action best describes a set of experimental data. The latter are single-photon probabilities $\tilde{\mathcal{P}}_{i,j}$, describing the transition from input mode $i$ to output mode $j$ (Fig. \ref{fig:schematriangolo}b), and Hong-Ou-Mandel \cite{HOM87} visibilities $\tilde{\mathcal{V}}_{ij,pq}$, describing two-photon interference from input modes $(i,j)$ to output modes $(p,q)$ (Fig. \ref{fig:schematriangolo}c). The associated errors are $\Delta \tilde{\mathcal{P}}_{i,j}$ and $\Delta \tilde{\mathcal{V}}_{ij,pq}$, respectively. Hong-Ou-Mandel visibilities are defined as $\mathcal{V}_{ij,pq} = (\mathcal{P}^{\mathrm{d}}_{ij,pq}-\mathcal{P}^{\mathrm{q}}_{ij,pq})/\mathcal{P}^{\mathrm{d}}_{ij,pq}$, where $\mathcal{P}^{\mathrm{d}}_{ij,pq}$ is the probability for two distinguishable particles and $\mathcal{P}^{\mathrm{d}}_{ij,pq}$ is the probability for two indistinguishable photons. The visibilities can be measured experimentally by recording the input-output coincidence pattern as a function of the relative delay $\Delta \tau$ between the input photons.  We model the initial group of unitaries by the set $\Phi = \{ E_1, ..., E_N \}$ of $N$ randomly-chosen individuals $E_l$. Every individual $E_l$ is completely determined by a set of real parameters, that represent its DNA. At a first glance, one could consider the elements of the unitary transformation (moduli and phases) to compose the DNA of the individuals. However, this is not the most appropriate choice since the generation of new offsprings from the random recombination of the parents according to this mechanism can lead to a non-unitary matrix. A better approach is obtained by exploiting the result by Reck et al. \cite{Reck94}, which showed that it is possible to decompose any linear $m \times m$ transformation in a network composed of phase shifters (PS) and beam splitters (BS) (see Fig. \ref{fig:schematriangolo}a). Every $k$-th PS-PS-BS set is defined by the transmittivity $t_k \in [0,1)$, and by the phases $\alpha_k, \beta_k \in [0,\pi]$. The DNA of the individual $E_l$ is then represented by the vector $E_l=  \{ G^l_1, G^l_2, ..., G^l_{M} \},$ with the parameter triples $G^l_k = \{ t^l_k, \alpha^l_k, \beta^l_k \}$ being the genes and $M=\sum_{g=1}^{m-1} g$ the total number of PS-PS-BS sets. The global unitary of the system $U$ can be obtained by multiplying the set of $m \times m$ unitary matrices $\mathcal{U}_k^{(m)}$ describing the action of the $k$-th gene. Each matrix is obtained starting from the $m \times m$ identity and replacing the elements corresponding to the involved modes with the ones of a PS-PS-BS $2 \times 2$ matrix. With such a parametrization, the unitariety of overall transformation is naturally guaranteed. This decomposition does not necessarily represent the actual internal structure of the system being studied, which may be in general unknown. Indeed, it represents a mathematical tool to reduce a unitary matrix as the combination of its independent unitary parts, those corresponding to the genes of the DNA.

The genetic algorithm \cite{SuppInfo} requires the definition of three ingredients governing the genetic evolution: (i) the \textit{fitness function}, which quantifies the survival probability of a given individual, (ii) the \textit{crossover function}, which governs the reproduction mechanism and (iii) the \textit{mutation process}, which enlarges the available set of genes and thus increases the variabililty of the individuals. The fitness function $f(E) \in \mathbb{R}:[0,\infty)$ is related to the survival probability of the individual $E$. In our case $f(E)$ is chosen to be inversely proportional to the distance between experimental data and the data generated from the unitary matrix $\mathcal{U}_{E}$ corresponding to the individual $E$. Assuming the matrices $\mathcal{P}_{i,j}^{E}$ and $\mathcal{V}_{ij,pq}^{E}$ to be, respectively, one-photon and two-photons measurement predictions generated from $\mathcal{U}_E$, we define the fitness function as $f(E) = 1/\chi^{2}$, where $\chi^{2}=\chi^{2}_{\mathcal{P}} + \chi^{2}_{\mathcal{V}}$ is the chi-square function composed by the two terms
\begin{equation}
\label{eq:chi_square}
\chi^{2}_{\mathcal{P}}= \sum_{i,j} \frac{(\tilde{\mathcal{P}}_{i,j}-\mathcal{P}_{i,j}^{E})^{2}}{\Delta \tilde{\mathcal{P}}^{2}_{i,j}}; \chi^{2}_{\mathcal{V}} = \sum_{i,j,p,q} \frac{(\tilde{\mathcal{V}}_{ij,pq}-\mathcal{V}_{ij,pq}^{E})^{2}}{\Delta \tilde{\mathcal{V}}^{2}_{ij,pq}}.
\end{equation}
In other words, the fitness $f(E)$ represents the quality of the solution $E$ for the given problem. The crossover function corresponds to the reproduction mechanism described above. Two individuals $E_A$ and $E_B$ generate one child $E_C$ whose DNA is composed of half genes from parent $E_A$ and the other half from parent $E_B$, randomly chosen. In the crossover mechanism, even if genes are randomly chosen they always occupy the same place in the child's DNA sequence. This means that PS-PS-BS sets inherited by the parents always occupy the same position within offsprings' corresponding linear optical networks. Finally, we establish that for any iteration of the algorithm any gene $G_k^l$ has a probability $\gamma$ (called mutation rate) of being replaced by a new random triple $\{ \tilde{t}^l_k, \tilde{\alpha}^l_k, \tilde{\beta}^l_k \}$. This probability must be carefully chosen. Indeed, an exceedingly high mutation frequency would reduce the search process to a random walk in the space of solutions, while an extremely small value would prevent the algorithm to reach the global maximum of $f(E)$. One of the key aspects of this algorithm is its computational efficiency. The price to pay is the reduction of the system governability, since the algorithm evolution is not deterministic. Furthermore, the optimal combination of parameters (mutation rate, population size, etc.) cannot be derived a priori and may depend on the dimension $m$ of the network.

\begin{figure}[b!]
\centering
\includegraphics[width=0.49\textwidth]{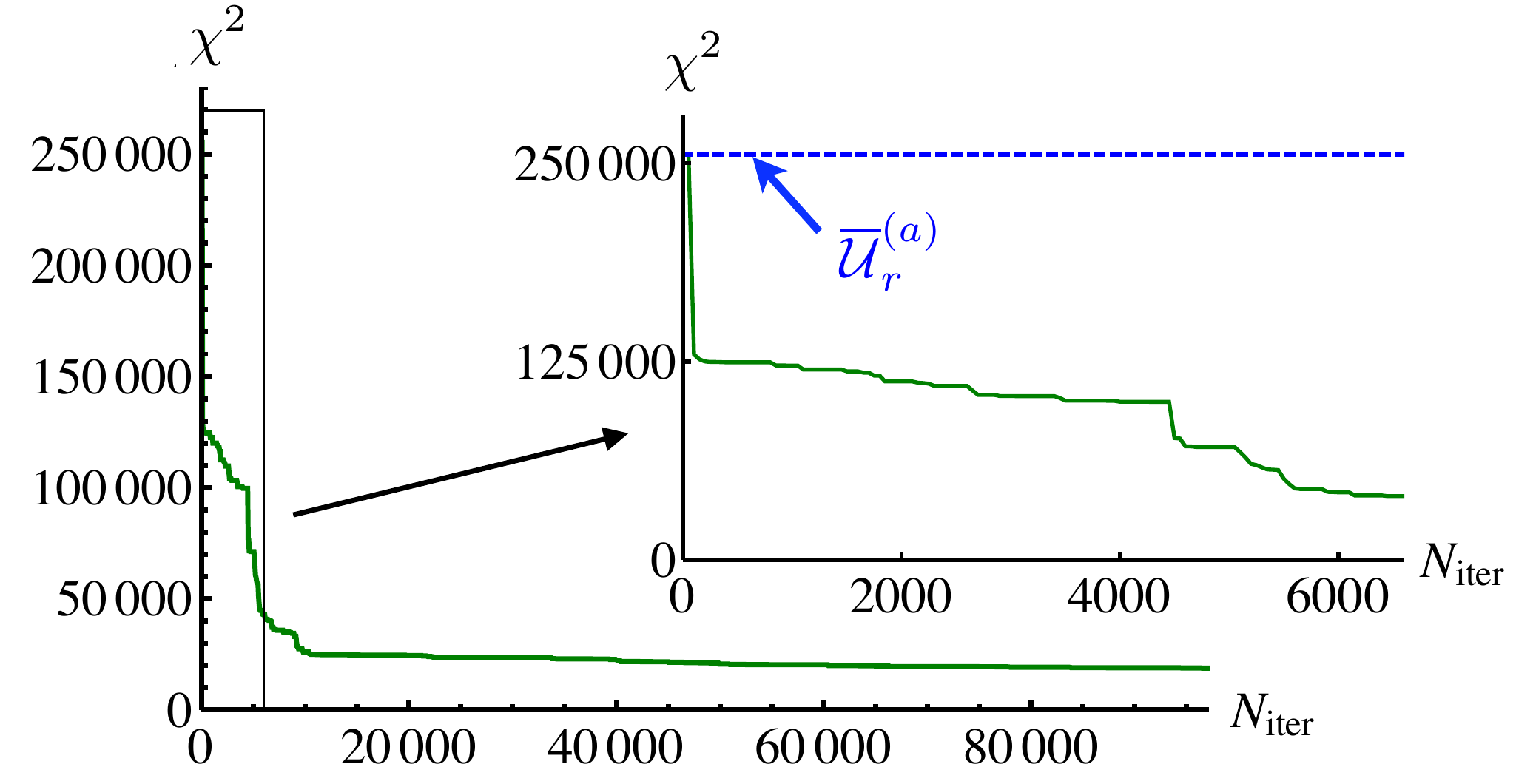}
\caption{Evolution of the minimum $\chi^{2}$ in the genetic pool with respect to the experimental data through the running time of the genetic algorithm, plotted as a function of the number of iterations. Green solid lines: minimum $\chi^{2}$ in the genetic pool. Inset: highlight for $N_{\mathrm{iter}} \in [0;6600]$, showing the action of mutations (jumps) and crossover (smooth variations). Horizontal blue dashed line: best $\chi^{2}$ obtained from the analytic method, which acts as a staring point of the genetic algorithm.
}
\label{fig:reco_genetic_convergence}
\end{figure}

\textit{Experimental results.-} We tested the genetic algorithm by reconstructing the linear transformation induced by a $7$-mode integrated interferometer, fabricated in a borosilicate glass substrate by means of the femtosecond laser waveguide writing \cite{gattass2008flm,dellavalle2009mpd} technique. This approach exploits the permanent and localized increase in the refraction index obtained by nonlinear absorption of focused femtosecond pulses, thus directly writing waveguides in the material. The internal structure of the implemented interferometer is shown in Fig. \ref{fig:schematriangolo}d, and is composed by a network of symmetric $50-50$ directional couplers and a phase pattern. We observe that the internal structure of the interferometer is different from the triangular structure adopted in the genetic algorithm to decompose the unitary matrices. Indeed, the adoption of the latter choice in the reconstruction algorithm is only a mathematical tool, which may not correspond to the actual structure of the transformation under analysis. Single-photon and two-photon input states, necessary to measure the data set of the algorithm, were prepared by a spontaneous parametric down conversion source and injected into the different input ports of the interferometer (see Supplementary Material \cite{SuppInfo}). The interference pattern necessary to measure the visibilities $\tilde{\mathcal{V}}_{ij,pq}$ was obtained by a controlling photon indistinguishability through a variable temporal delay between the input particles.

The reconstruction method based on the genetic approach has been applied to the $7$-mode chip. The complete set of experimental measurements consists of $d_{1}=49$ single photon transition probabilities $\tilde{\mathcal{P}}_{i,j}$ and $d_{2}=441$ two-photon Hong-Ou-Mandel visibilities $\tilde{\mathcal{V}}_{ij,pq}$, thus corresponding to an overcomplete set of $d=d_{1}+d_{2}=490$ experimental data. The genetic algorithm maximizes the fitness function $f(E)$ [Eq. (\ref{eq:chi_square})] between the experimental data set and the predictions $\mathcal{P}_{i,j}^{E_{l}}$ and $\mathcal{V}_{ij,pq}^{E_{l}}$ obtained from the unitary $\mathcal{U}_{E_{l}}$ belonging to the population of the genetic algorithm. The starting point of the protocol is a population of $s=100$ unitaries. As a modification to the recipe previously discussed, a subset of $s_{1}=20$ unitaries at the initial step is chosen starting from the algorithm introduced in Ref. \cite{Obrien12}. With that method, a minimal set of single- and two-photon data is exploited to retrieve analytically the elements of the unitary matrix. This approach can be extended by considering that a set of $m^2$ independent estimates of $U$ can be obtained by recording the full set of single- and two-photon measurements, and by permuting the mode indexes accordingly \cite{Crespi2012}. These operations correspond to selecting $m^2$ independent minimal data sets. For the genetic algorithm, we then choose the $s_{1}=20$ unitaries (among the set of 49 possible matrices for $m=7$) presenting the lower values of the $\chi^{2}$ with respect to the full set of experimental data, that is, having higher fitnesses. This provides a reasonable starting point for the genetic pool. Finally, the remaining subset of $s_{2}=80$ are randomly generated from the Haar measure. 

In Fig. \ref{fig:reco_genetic_convergence} we report the evolution of the best $\chi^{2}$ in the genetic pool during the running time of the algorithm. We observe that an almost stable value of the $\chi^{2}$ is obtained after $N_{\mathrm{iter}} \sim 40000$ iterations, corresponding to a computational time of $t \sim 1$ h on a laptop. During the evolution, the decrease of the $\chi^{2}$ (the increase of the fitness) occurs with two different trends (see inset of Fig. \ref{fig:reco_genetic_convergence}). Smooth variations are due to the crossover mechanism between members of the population, converging to the best possible unitary given the available genetic pool. Conversely, fast jumps in the $\chi^{2}$ are due to random mutations in the genetic pool. The convergence of the genetic algorithm is confirmed by the decrease of the chi square $\chi^{2}$ from the starting value $\min_{\mathcal{U}_{r}^{(a)}} \chi^{2} \sim 255000$, obtained from the best unitary $\mathcal{\overline{U}}_{r}^{(a)}$ of the analytic approach, to a final value of $\chi_{r,(g)}^{2} \sim 17096$, leading to an improvement of one order of magnitude. As an additional figure of merit, we consider the similarities $S_{r}^{(a)}$ between the experimental two-photon visibilities and the predictions obtained from the analytic unitaries $\mathcal{U}_{r}^{(a)}$, according to the definition $S_{r}^{(a)} = 1- \sum_{i,j,p,q} \vert \tilde{\mathcal{V}}_{ij,pq} - \mathcal{V}_{ij,pq}^{r,(a)}\vert/(2 d_{2})$ (and analogous definition for the output of the genetic approach). We observe that the similarity $S_{r}^{(g)}$ obtained for the output unitary $\mathcal{U}_{r}^{(g)}$ from the genetic algorithm, equal to $S_{r}^{(g)} = 0.957 \pm 0.001$, clearly outperforms the maximum value obtained from the analytic algorithm: $\max_{\mathcal{U}_{r}^{(a)}} S_{r}^{(a)} = 0.920 \pm 0.001$.

\begin{figure}[ht!]
\centering
\includegraphics[width=0.49\textwidth]{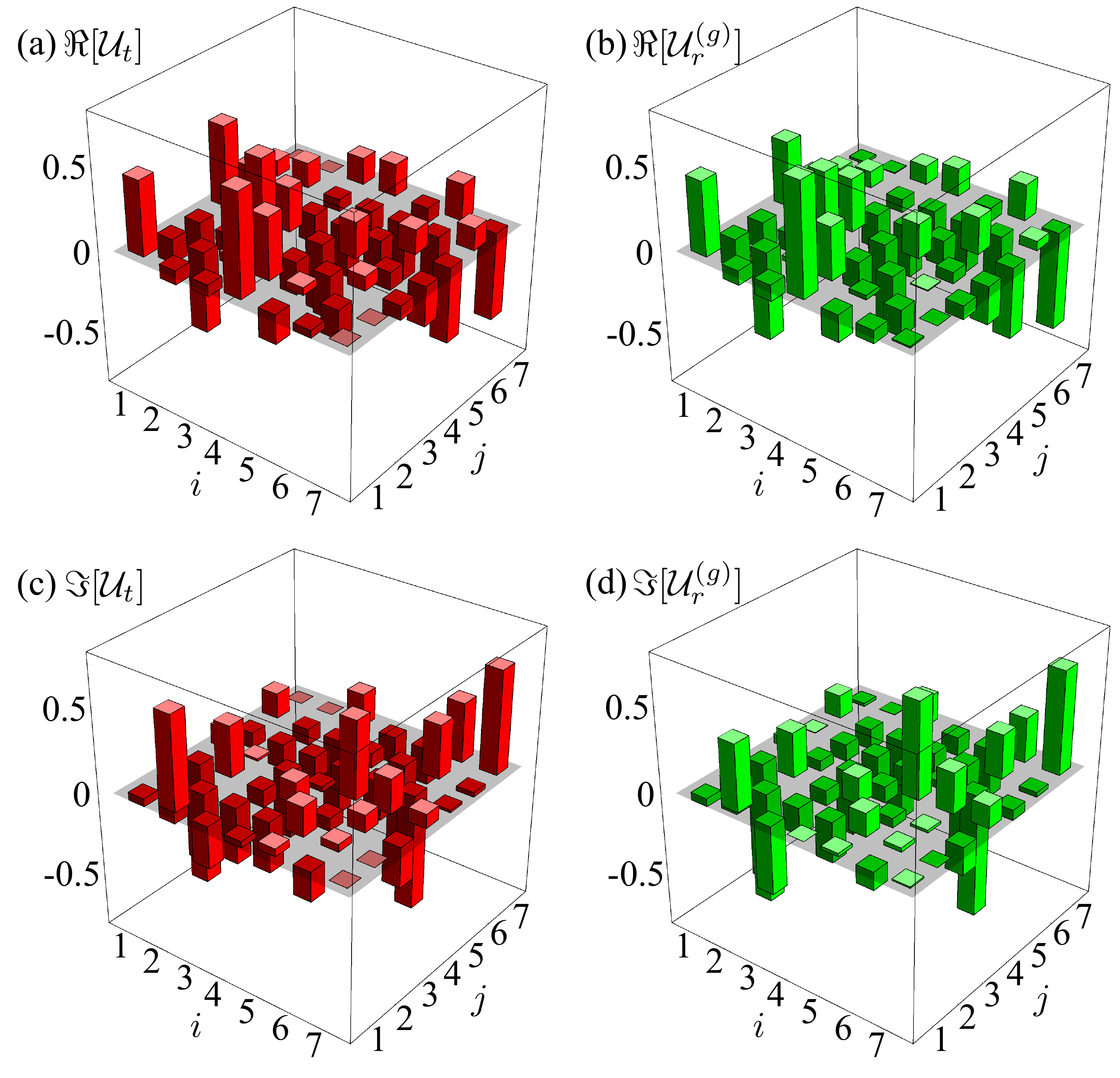}
\caption{(a), Real part of the theoretical unitary matrix $\mathcal{U}_{t}$. (b), Real part of the reconstructed unitary matrix $\mathcal{U}_{r}^{(g)}$. (c), Imaginary part of the theoretical unitary matrix $\mathcal{U}_{t}$. (d), Imaginary part of the reconstructed unitary matrix $\mathcal{U}_{r}^{(g)}$.}
\label{fig:reco_genetic_unitary}
\end{figure}

The results for the obtained unitary are shown in Fig. \ref{fig:reco_genetic_unitary}, where the real and imaginary parts of the output unitary matrix of the genetic algorithm $\mathcal{U}_{r}^{(g)}$ are compared with the theoretical unitary $\mathcal{U}_{t}$ expected from the fabrication process. The gate fidelity between $\mathcal{U}_{t}$ and $\mathcal{U}_{r}^{(g)}$, defined as $F_{r,t}^{(g)} = \vert \mathrm{Tr}[\mathcal{U}_{t}^{\dag} \mathcal{U}_{r}^{(g)}] \vert/m$, reaches a value $F_{r,t}^{(g)} = 0.975 \pm 0.013$. This parameter represents the quality of the implemented unitary in the fabrication process indicating how close the implemented interferometer is with respect to the ideal one. The error on the gate fidelity has been estimated by a $N=10000$ Monte-Carlo simulation of unitary reconstruction with the analytic method.

Further optimizations of the protocol can be envisaged. For instance, the $\chi^{2}$ function in the fitness may be replaced with a weighted function $\chi^{2}_{\mathrm{w}} = w \chi^{2}_{\mathcal{P}} + (1-w) \chi^{2}_{\mathcal{V}}$. We then performed the reconstruction method for different values of the weight $w$, observing that for the present data the best choice is obtained for the symmetric case $w=0.5$. The optimal weight $w$ can nevertheless vary with the dimension $m$ of the network. Additionally, the number of unitaries $s_1$ taken at the initial step from the analytic algorithm can be optimized depending on the problem size.

\textit{Conclusions and perspectives.-} In this letter we have described an approach to learn an unknown linear optical process $\mathcal{U}$ by exploiting a specifically tailored genetic algorithm. We have then tested the present approach for the reconstruction of an unknown $7 \times 7$ integrated linear optical interferometer built by the femtosecond laser-writing technique. The experimental results show that this methodology is suitable to be exploited for the characterization of linear optical networks with progressively increasing number of modes, with applications in different contexts such as quantum simulation and quantum interferometry. Further perspectives can be envisaged by applying these genetic approaches in the context of learning unknown patterns \cite{Schu15} or general Hamiltonian evolutions \cite{Gran12}. The algorithmic approach itself may be adapted so as to progressively change the parameters of its evolution or the measured data set sequence depending on the results of the previous steps.

\textit{Acknowledgments.-} We acknowledge very useful discussions with D. J. Brod and E. F. Galv\~ao. This work was supported by the ERC-Starting Grant 3D-QUEST (3D-Quantum Integrated Optical Simulation; grant agreement no. 307783): http://www.3dquest.eu, and by the H2020-FETPROACT-2014 Grant QUCHIP (Quantum Simulation on a Photonic Chip; grant agreement no. 641039): http://www.quchip.eu.


%

\end{document}